# Bidirectional elastic diode with frequency-preserved nonreciprocity

Xin Fang[1,2], Jihong Wen[1,*], Li Cheng[2], and Baowen Li[3,†]

[1]*Laboratory of Science and Technology on Integrated Logistics Support, College of Intelligent Science, National University of Defense Technology, Changsha, Hunan 410073, China*
[2]*Department of Mechanical Engineering, Hong Kong Polytechnic University, Hong Kong, China.*
[3]*Paul M Rady Department of Mechanical Engineering and Department of Physics, University of Colorado, Boulder, Colorado 80309, USA*

The study of nonreciprocal wave propagation is of great interests for both fundamental research and engineering applications. Here we demonstrate theoretically and experimentally a bidirectional, nonreciprocal, and high-quality diode that can rectify elastic waves in both forward and backward directions in an elastic metamaterial designed to exhibit enhanced nonlinearity of resonances. This diode can preserve or vary frequency, rectify low-frequency long wave with small system size, offer high-quality insulation, can be modulated by amplitude, and break reciprocity of both the total energy and fundamental wave. We report three mechanisms to break reciprocity: the amplitude-dependent bandgap combining interface reflection, chaotic response combining linear bandgap, amplitude-dependent attenuation rate in damping diode. The bidirectional diode paves ways for mutually controlling information/energy transport between two sources, which can be used as new wave insulators.

## I. Introduction

Diode, a one-way nonreciprocal road for energy transport, plays a pivotal role in electronics [1, 2]. Recent years have witnessed the extension of the concept to heat conduction [3-5] and elastic wave manipulation [6-8]. Major concerns of an elastic diode pertaining to practical applications include the frequency range, frequency variation, size/wavelength ($\lambda$) ratio, and insulation quality characterized by the transmission difference $\Delta T=|T_F-T_B|$, where $T_F$ ($T_B$) is the forward (backward) transmission. Several mechanisms have been proposed to realize nonreciprocal elastic wave propagation [8-10].

Topological states offer nonreciprocity in linear elastic systems [11-16] by beating time reversibility with active control or external field, while it is challenging to realize this mechanism in the broad and low frequencies [17-19]. Moreover, this manipulation relies on active control or external field.

Nonlinearity has been demonstrated to offer robust diode effect in two ways: harmonic generation and bifurcation [7, 20-23]. In an acoustic diode that consists of a segment of nonlinear medium and a segment of linear phononic crystal, high-order harmonics are generated in the nonlinear medium and the bandgap of the phononic crystal entails asymmetric transfer of the total energy [7, 24]. However, this diode [7] causes a frequency change in the output signal as compared with the input one [25], which results in information loss. Moreover, the thickness of this diode [7] is typically ~30$\lambda$, a very large size for low-frequency sound or vibration. The bifurcation- induced elastic diode can realize the frequency-preserved effect, while experimental realizations based on granular chains require a thickness of ~10$\lambda$ [20, 26, 27]. Actively shifting the bandgap of a time-variation metamaterial can also induce nonreciprocity [28-31], but the insulation is limited to $\Delta T \approx 7$ dB with a size of ~5$\lambda$ [32]. Moreover, studies should clearly distinguish nonreciprocity of the total energy and fundamental wave.

An open and crucial question is how to design an integrated high-quality, low-frequency, frequency-preserved and small elastic diode. The critical component to solve this problem is the provision of resonators with enhanced nonlinearity and a mechanism for frequency- preservation. Moreover, although bidirectional diodes are readily available in electronics, existing elastic diodes are only unidirectional, namely, they allow energy/information flow only in one direction.

*wenjihong@vip.sina.com;
†Baowen.Li@Colorado.Edu

This paper shall demonstrate both theoretically and experimentally a new type of elastic diode – frequency-preserved bidirectional diode. The reciprocal transports of total energy and fundamental wave are broken in both directions. Furthermore, this high-quality diode allows for low-frequency wave rectification with a thickness of ~λ. The achieved bidirectional diode opens new horizons for broad engineering applications as well as the exploration of new physics.

## II. Model and Methods

Our bidirectional diode consists of one segment of linear elastic metamaterial (LEM) [33, 34] and one segment of nonlinear elastic metamaterial (NEM) with enhanced nonlinearity, as illustrated in Fig. 1. This NEM offers many fascinating features [35, 36] including generating a low-frequency, self-modulated, amplitude-dependent nonlinear locally resonant (NLR) bandgap and chaotic responses. The sophisticated triatomic NEM reported in Ref. [37] is adopted to realize the critical mechanisms and it contains 10 metacells.

The forward direction is the case where the incident wave arriving from the left terminal first enters the LEM (Fig. 1(a)). Conversely, the backward corresponds to the case when the input wave first enters the NEM. In such a metacell (Fig. 1(c)), the primary oscillator $m_0$ is a hollowed parallelepiped, and two neighbouring $m_0$ units are coupled through a pair of springs whose entire stiffness is $k_0$. The hollow cylinder $m_2$ held inside $m_0$ has a hole whose radius is 4.045 mm. $m_2$ couples to $m_0$ by two springs with total stiffness $k_2$. A steel sphere $m_1$ is held at the centre point of the cylindroid cavity in $m_2$, and the two are connected by a curved spring whose stiffness is $k_1$. A tiny clearance $\delta_0$=45±15 μm between $m_1$ and $m_2$ is tactically created to generate enhanced nonlinearity through vibration-collision. As detailed in Appendix A, the nonlinear stiffness coefficient reaches $k_N$>3×10$^{12}$ N/m$^3$ if we equivalent this clearance nonlinearity to a smooth cubic nonlinearity.

The equations of motion for the individual masses of the $n^{th}$ triatomic cell are:

$$\begin{cases} m_0\ddot{u}_n = k_0(u_{n+1}+u_{n-1}-2u_n)+ \\ \quad c_0k_0(\dot{u}_{n+1}+\dot{u}_{n-1}-2\dot{u}_n)+k_2(z_n-u_n) \\ m_1\ddot{y}_n = -k_1(y_n-z_n)-F_N(t) \\ m_2\ddot{z}_n = -k_2(z_n-u_n)+k_1(y_n-z_n)+F_N(t) \end{cases} \quad (1)$$

Here, $u_n$, $y_n$, and $z_n$ denote the barycenter displacements of $m_0$, $m_1$, and $m_2$ in the $n^{th}$ cell, respectively; $F_N(t)$ denotes the nonlinear contact force depending on $\delta_0$. Unless otherwise specified, a weak viscous damping, $c_0k_0$, $c_0$=3×10$^{-5}$ s, in the primary chain is taken into consideration in theoretical models. To grasp the general physics, we use a smooth cubic nonlinearity in simulations, $F_N(t)=k_N(y_n-z_n)^3$, and specify $k_N$=3×10$^{12}$ N/m$^3$.

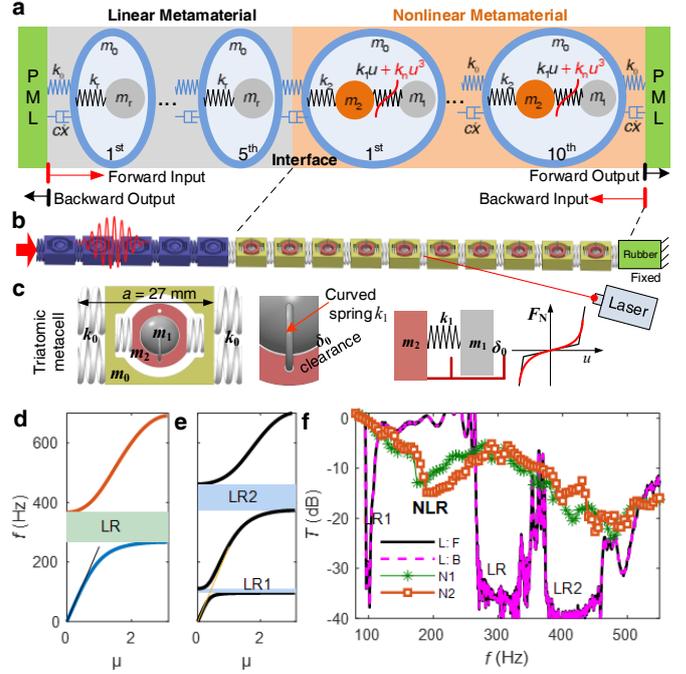

FIG. 1 (Color online) Elastic diode. (a) Schematic diagram of our bidirectional diode consisting the LEM (left) and NEM (right). (b) Experimental setup of the diode. Only the forward propagation is indicated and the backward is similar. (c) Structure of the nonlinear triatomic metacell and its equivalent mathematical model. (d) and (e) Dispersion curves of the linearized diatomic and triatomic EMs, respectively. (f) Numerical transmissions: Here, "L: F" and "L: B" denote the forward and backward transmission of the linearized model, respectively. "N1" and "N2" correspond to transmissions only in the NEM (without LEM) for the incident amplitude $A_0$=40 and 60 μm, respectively, where the amplitude dependence is seen. In (f), the damping coefficient $c_0$ = 3×10$^{-5}$ s.

The LEM consists of 5 linear diatomic metacells whose primary structures are identical with that of the triatomic cell. We fix the sphere into the cavity of the cylinder to form the local resonator $m_r$, i.e., no clearance here. Therefore, the equations of motion for the $n^{th}$ diatomic metacell are:

$$\begin{cases} m_0\ddot{u}_n = k_0(u_{n+1}+u_{n-1}-2u_n)+ \\ \quad c_0k_0(\dot{u}_{n+1}+\dot{u}_{n-1}-2\dot{u}_n)+k_2(r_n-u_n) \\ m_r\ddot{r}_n = -k_r(r_n-u_n), \quad k_r=k_2 \end{cases} \quad (2)$$

where $r_n$ denotes the barycenter displacements of $m_r$. The parameters are: masses $m_0$=5.8, $m_1$=2.1, $m_2$=2, $m_r$=3.72 g; $k_r=k_2$; $\omega_i=2\pi f_i=\sqrt{k_i/m_i}$, $i$=0, 1, 2, $r$, and $f_0$=322, $f_1$=100, $f_2$=390.6, $f_r$=286.4 Hz.

Moreover, we adopt the equivalent approach based on the bifurcation of nonlinear resonance to analytically explain the dispersion property of the NEM [37] (see Appendix B). The dispersion equation describes the relationship between the wave vector $\kappa$ and frequency $f$. $\omega=2\pi f$. $\mu=\kappa a=\mu_R+i\mu_I$, $a$=27 mm denotes the lattice constant. Due to the amplitude attenuation, $\mu_I$ self-adaptively changes with propagation distance, $n$. The attenuation from $(n-1)^{th}$ to $n^{th}$ metacell is $A(n)/A(n-1)= \exp[-\mu_I(n)]$.

There are two types of incident signals: (1) A 10-period sinusoidal packet described by the function $u_0=A_0\sin(\omega t)[1-\cos(\omega t/10)]/2$, $\omega=2\pi f$; (2) The standard sinusoidal wave, $u_0=A_0\sin(\omega t)$. Numerical integration method is used to solve the wave propagation in models. We confirm that all simulation processes are convergent. Numerical errors are smaller for inputting the wave packet (than inputting the standard sinusoidal wave). Therefore, we use the wave packet as the incident signal in most cases to obtain the distribution of transmission $T(f, A_0)$. The standard sinusoidal wave is input when we need to analyse the frequency spectra. In fact, regularities indicated by the two types of signals are same.

Moreover, to eliminate any asymmetric transmissions due to boundary conditions, we calibrate the numerical diode model whose both ends are connected to a long perfect match layer (PML) consisting of 120 unit cells (see Supplementary material).

We distinguish the diode effects in time and frequency domains. The time-domain wave amplitude is the peak value of $u_n(t)$, $A_n^{(time)}$=Peak$[u_n(t)]$. The frequency-domain amplitude $A_n^{(freq)}(f)$ is picked from the spectrum of $u_n(t)$ at the incident frequency. Transmissions in time and frequency domains are

$$T_{time}(f, A_0) = 20\log_{10}[A_{out}^{(time)}(f)/A_0(f)] \quad (dB) \quad (3)$$

$$T_{freq}(f, A_0) = 20\log_{10}[A_{out}^{(freq)}(f)/A_0(f)] \quad (dB) \quad (4)$$

where $A_{out}(f)$ is the output wave amplitude. The output port is the 15$^{th}$ $m_0$ from the exciter. As there are multiple harmonics, $T_{time}$ represents the transmission of total energy containing all frequency components; $T_{freq}$ represents the transmission of fundamental wave. $\Delta T=T_F-T_B$.

## III. Bidirectional rectification with preserved frequency

As shown in Fig. 1(d-f), the diatomic LEM has a local resonant (LR) bandgap within 267-366 Hz, and the linearized triatomic EM exhibits two LR bandgaps: LR1 96-110 Hz, LR2 374-462 Hz. The NEM produces an NLR bandgap that depends on the incident amplitude $A_0$. For example, for $A_0$=40 μm, NLR appears in 210-275 Hz that is in the passbands of the diatomic and triatomic LEM [37]. The slope of the dispersion curve for $f\rightarrow 0$ indicates that the wave speed of the long wave is $v_0$=1000$\pi a$. In Fig. 1(f), the perfectly symmetric transmission of linear structure also validates the numerical model including the PML, and all three bandgaps, LR, LR1 and LR2, are activated to produce significant wave attenuation due to the large attenuation $\mu_I$ (see Fig. 3(a)).

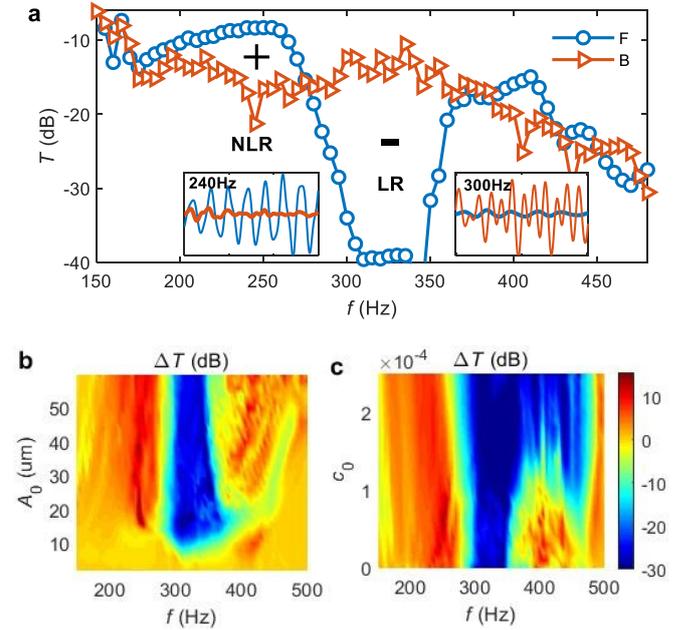

FIG. 2 (Color online) Numerical results of bidirectional diode. All transmissions denote the time-domain value $T_{time}$. (a) Forward (F) transmission $T_F$ and backward (B) transmission $T_B$. Here, the inserted panels are waveforms of the forward and backward transmitted waves for 240 Hz and 300 Hz. (c) and (d) Distributions of $\Delta T$ versus frequency and $A_0$ (or $c_0$), respectively. Unless otherwise specified, $A_0$=40 um and $c_0$=3×10$^{-5}$ s

Firstly, we consider $T_{time}$. In the diode (Fig. 2(a)), both $T_F$ and $T_B$ show attenuations for $f$>170 Hz due to enhanced nonlinearity and self-adaptive bandgap effect [37], but the forward and backward reductions are different. $T_F$ reaches -40 dB in 290-350 Hz due to the LR bandgap. $T_F \approx$ -10 dB due to the NLR bandgap near 250 Hz. By contrast, the backward waves achieve greater reductions ($T_B \approx$-20 dB) in the NLR range, but its attenuation in the LR bandgap ($T_B \approx$-15 dB) is much smaller than $T_F$. These observations suggest that our integrated metamaterial breaks reciprocity in both NLR and LR bandgaps, and surprisingly, their nonreciprocal directions are opposite, which means that the diode can rectify wave energy in both directions for different frequencies. We call this new phenomenon *dual nonreciprocity* or *bidirectional diode*. As is demonstrated by $\Delta T(f, A_0)$ in Fig. 2(b,c), the dual nonreciprocity is enhanced by with increasing $A_0$ and it can appear without damping. However, increase the damping can optimize $\Delta T$ in the NLR range. Importantly, $\Delta T$ reaches 15 dB and -30 dB in NLR and LR ranges, respectively, thus featuring high-quality diode effect. Moreover, the wavelength in the nonreciprocal band is of $2.7\pi a<\lambda<5\pi a$, which means the thickness of our diode, $15a$, is only of $\sim\lambda$. Therefore, a small diode is achieved to rectify low-frequency waves by using the sub-wavelength metacell.

## IV. Physical Mechanisms and properties of the diode

In order to understand the underlying physics and properties of the diode, we analytically study the amplitude-dependent dispersion relationship $\mu_I(f)$ of the NEM (Fig. 3). In nonlinear cases, the band near the peak $\mu_I$ is the NLR bandgap shifted from LR1 shown in Fig. 3(a) [37]. The blue shading region 1 in Fig. 3(a, b) denotes the difference of the attenuation rate $\Delta\mu_I$ in NLR range, induced by amplitude. The green shading region 2 is $\Delta\mu_I$ in the passband between the strongly and weakly nonlinear (or linear) metamaterial. In theory, nonreciprocity of fundamental wave is described by $\Delta\mu_I$ between the forward and backward processes in the NEM:

$$\Delta T_{\text{freq}} = 20\log_{10}\left\{\exp\left[-\sum_{n=1}^{10}\Delta\mu_I(n)\right]\right\} \quad \text{(dB)} \qquad (5)$$

In region 2, $\mu_I=0$ for $c_0=0$ (Fig. 3(a)), which means the amplitude has no influence on nonreciprocity in this case.

We performed simulations at typical frequencies to show these nonreciprocal processes. As shown in Fig. 4, the transmissions $\Delta T_{\text{time}}$ and $\Delta T_{\text{freq}}$ at $n^{\text{th}}$ cell from the incident source, there are some differences between by inputting the aforementioned wave packet and monochromous wave. As detailly analyzed in the supplementary material, those differences arise from two aspects: (1) The wave packet is a continuous-spectrum signal while the standard sinusoidal wave is monochromous; (2) Inputting the monochromous wave may induce standing-wave resonance because of the imperfections of the "perfectly match layer" (i.e., the PML cannot absorb 100% incident energy). Fortunately, properties indicated by the two types of incident signals are same.

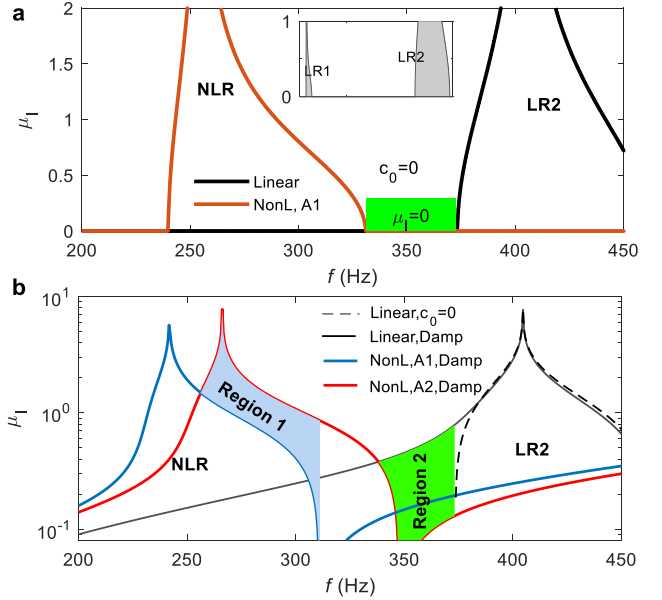

FIG. 3 (Color online) Mechanisms for nonreciprocity. (a) Analytical dispersion relationship $\mu_I(f)$ of the undamped metamaterial, $c_0=0$. (b) $\mu_I(f)$ of the dappled metamaterial. Here, $c_0=2\times10^{-4}$ s in the damped cases; A1 ($A_0=30$ μm) and A2 ($A_0=60$ μm) represent the moderate and high amplitudes in nonlinear model, respectively.

### A. Amplitude-dependent bandgap

As illustrated in Fig. 4(a), for waves in the NLR bandgap represented by 240 Hz, although the spectrum of the forward transmitted wave is complex and the backward propagation features second-harmonic generation, fundamental wave remains the main component, which highlights a *frequency-preserved single-mode* operation. The varying trends of $T_{\text{time}}$ and $T_{\text{freq}}$ in Fig. 4(b) clarify that reciprocity of both the total energy and fundamental wave are broken, and $\Delta T_{\text{freq}}$ is much higher.

The asymmetric transmission in the NLR bandgap originates from the amplitude-dependent property. In the NEM, the initial amplitude in the backward direction is $A_0$. At the interface between LEM and NEM, the forward incident wave is partly reflected by NEM due to its NLR bandgap (see Fig. 4(b,d)). Therefore, the initial amplitude of the transmitted wave in NEM is smaller than $A_0$. In essential, the amplitude difference between forward and backward directions forms the shading region 1 in Fig. 3(b). According to Eq. (5), nonreciprocity in this range is realized because the attenuation factor $\Delta\mu_I\neq0$ between two directions. The backward wave obtains higher attenuation, namely, larger $\mu_I$ due to higher amplitude.

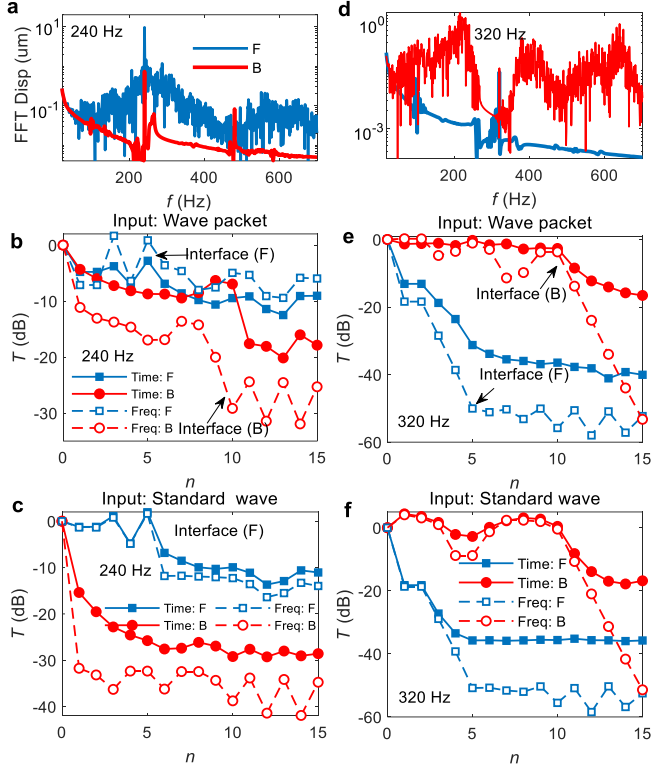

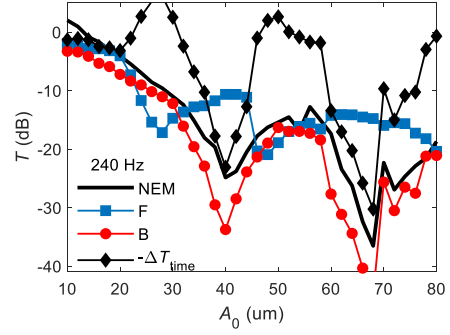

FIG. 4 (Color online) Properties of the bidirectional diode. $A_0$=40 μm, $c_0$=3×10$^{-5}$ s if it is constant. (a-c) for 240 Hz; (d-f) for 320 Hz. (a, d) Displacement spectra at the 15$^{th}$ cell in forward and backward directions. (b, e, c, f) Transmissions $\Delta T_{time}$ and $\Delta T_{freq}$ at $n^{th}$ cell from the incident source. Input waves in (b, e) are wave packets, and in (c, f) are monochromic sinusoidal wave signals.

Furthermore, by describing the offset between $T_F(A_0)$ and $T_B(A_0)$, we present a phenomenological explanation for this nonreciprocal propagation, as shown in Fig. 5. Although the entire nonreciprocity in this range is almost constant (see Fig. 2(a)), the amplitude-dependent $\Delta T_{time}$ in Fig. 5 fluctuates with $A_0$. The positive $\Delta T_{time}$ at 40 and 68 μm reaches 20 and 30 dB, respectively: high-quality diode. The transmission of the 10-cell NEM, $T_{NEM}$, decreases with increasing $A_0$ in whole, but local fluctuations lead to minimal $T_{NEM}$. Because $T_{NEM}$ and $T_B$ follow the same trends including the valleys: $T_{NEM} \propto T_B$. In contrast, at the interface between LEM and NEM, the forward incident wave is partly reflected by NEM due to its NLR bandgap. Therefore, the initial amplitude of the transmitted wave in NEM is smaller than $A_0$, so the curve of $T_F(A_0)$ is shifted to the upper left relative to $T_B(A_0)$: the offset between $T_F(A_0)$ and $T_B(A_0)$ likes the phase difference between two mathematical functions. Damping in the LAM offers the similar offset effect. As the valley of $T_B(A_0)$ meets the peak of $T_F(A_0)$, significant nonreciprocity is generated. Nevertheless, nonreciprocity diminishes in some amplitude ranges when the peak of $T_B(A_0)$ meets the valley of $T_F(A_0)$.

FIG. 5 (Color online) Transmission $T_{time}$ of NEM and $\Delta T$ of the diode change with $A_0$ under 240 Hz. To show laws more clearly, -$T$ is depicted here. Input waves are wave packets

### B. Chaotic effect

For waves in LR band, as represented by 320 Hz, the output frequency is changed (Fig. 4(d)). Because of LR bandgap, the forward wave is mostly reflected by LEM that leads to a reduction of 40 dB within 5 cells (see Fig. 4(e, f)). Upon entering the triatomic metamaterial, the amplitude of the wave (i.e., the residual energy) is already very small, which can just cause linear or very weak nonlinear effect. Therefore, this wave goes into the passband of the triatomic metamaterial that can go through it with little attenuation. As a result, $T_F^{time} \approx$ -40 dB, and the frequency remains in this process. However, for backward propagation, the same wave first entering the triatomic metamaterial generates enhanced nonlinearity that induces chaotic wave dynamics featuring continuous spectrum [38, 39], namely, the entering wave energy spreads over to other frequencies that can go through the LEM. Changes of $T_B^{time}$ with the propagation distance show that the attenuation mainly occurs in the LEM due to the LR bandgap. However, the reduction is only 15 dB because the LR cannot effectively reflect the energy with frequencies outside of this band. Consequently, this leads to a rectification of $|\Delta T_{time}|$>25 dB. Therefore, the nonreciprocity in LR is a *single-mode to chaotic continuous-mode* operation.

### C. Damping effect

However, $T_F^{freq} \approx T_B^{freq}$ at $n$=15 in Fig. 4(e, f) indicates that the chaotic mechanism only breaks the reciprocity of total energy while almost preserves reciprocity at fundamental frequency for weak damping. The mechanism for this "abnormal" weak nonreciprocity lies in region 2 in Fig. 3(a): $\Delta\mu_I$=0 for $c_0$=0, thus, $\Delta T_{freq}$=0 in Eq. (5). The weak nonreciprocity in this case originates from the energy transfer to other bands.

Interestingly, according to Eq. (5), by analytically calculating $\mu_I(n)$ at every metacell, we find that reciprocity of fundamental wave in LR band can be broken by increasing the damping, as shown in Fig. 6(a). Numerical results in Fig. 6(b-d) confirm this property. The secrete still lies in the shading region 2: $\Delta\mu_I \neq 0$ between the strongly nonlinear case (i.e., the backward propagation) and weakly nonlinear case (i.e., the forward propagation) if there is damping. Moreover, increasing damping can highlight the fundamental wave from the chaotic spectrum (Fig. 6(c)). The maximum $\Delta T_{freq}$ appears near 340 Hz in Fig. 6(b) because the valley value $\mu_I \rightarrow 0$ is here.

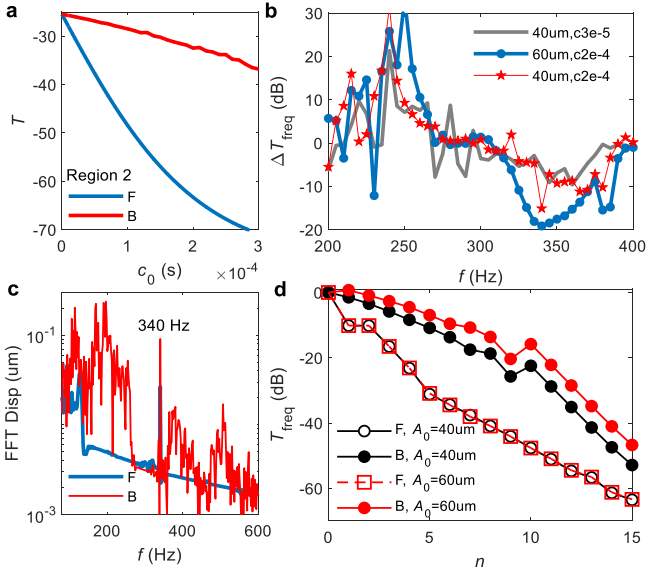

FIG. 6 (Color online) Wave propagation in the shading region 2 Here we take $f$=340 Hz as example. (a) Analytical transmissions of the diode when changing $c_0$, $A_0$=60 μm. (b) $T_{freq}$ for different damping cases. (c) Frequency spectrum for $A_0$=60 μm, $c_0$=2×10$^{-4}$ s. (d) Propagating processes for $A_0$=40 and 60 μm. Input waves are standard monochromic signals.

### D. Sub-wavelength diode

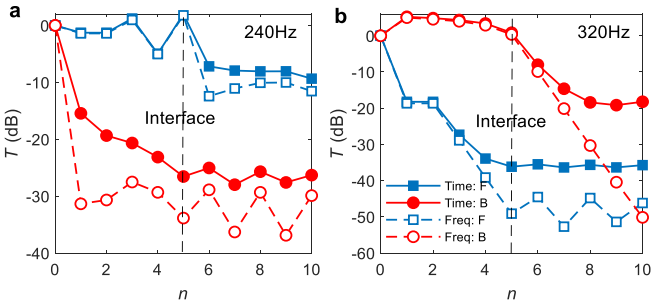

FIG. 7 (Color online) Nonreciprocal properties for the diode model consisting of 5 diatomic and 5 triatomic metacells. Input waves are standard sinusoidal waves. $A_0$=40 um and $c_0$=3×10$^{-5}$ s. (a) 240z. (b) 320 Hz.

The bidirectional nonreciprocity stated above are reported with the diode model consisting of 5 diatomic and 10 triatomic metacells (i.e., a "5+10" model). Actually, a shorter diode, e.g., the "5+5" model with ~0.67 wavelength, can also offer highly efficient nonreciprocal wave control, as shown in Fig. 7. The properties are same with the laws in Fig. 4(c, f).

### E. Wave interference

Moreover, as shown in Fig. 4(f) and Fig. 7(b), when inputting the standard sinusoidal wave at 320 Hz, the backward transmission at some positions $n$ slightly exceeds 1 (i.e., 4 dB). This is the interference between the incident wave and the reflected wave from the interface between the triatomic and diatomic segments. The evidences are shown in Fig. 8. Two models with different lengths "5+10" and "5+20" are analyzed here. In the backward direction (Fig. 8(a,b)), the wave amplitude in the linear triatomic segment experiences much greater fluctuation (especially near the interface) than that in the nonlinear model because nonlinearity reduces the interference. Whereas in forward direction (Fig. 8(a)), waves in both linear model (i.e., the model for $k_N$=0 in the triatomic segment) and nonlinear models experience the same fluctuation because the amplitude becomes so small that linearity dominates the dynamics.

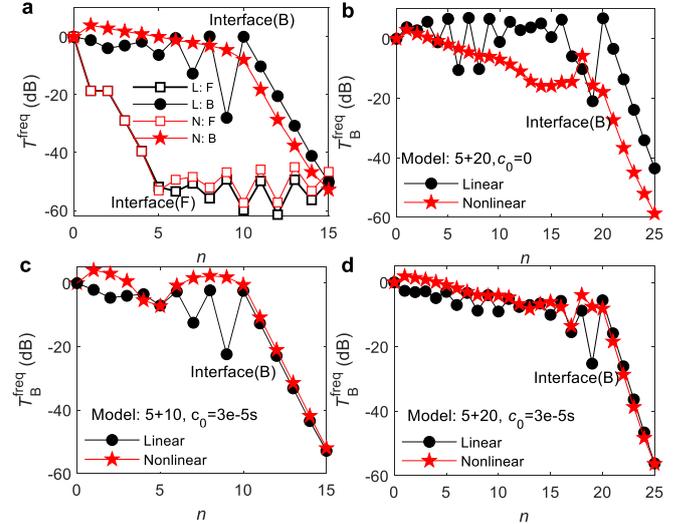

FIG. 8 (Color online) Wave interference in the diode at 320 Hz, $A_0$=40 um. (a, c) The model consisting of 5 diatomic and 10 triatomic metacells. (b, d) The model consisting of 5 diatomic and 20 triatomic metacells. (a, b) Undamped cases. (b, c) Cases for weak damping $c_0$=3×10$^{-5}$ s. Here, all signals are analyzed in frequency domain, and only the backward transmissions $T_B^{freq}$ are shown (b-d). Linear (L) and nonlinear (N) model denote the cases $k_N$=0 and $k_N$=3×10$^{12}$ N/m$^3$ in triatomic metacells, respectively.

Moreover, as shown in Fig. 8(c,d), damping can weaken the fluctuation in linear model, but it enhances the fluctuation in the "5+10" nonlinear model (see Fig. 8(c)) because damping can weaken the waveform distortion in certain extent. When increasing the length of the triatomic segment to 20 metacells (Fig. 8(d)), this phenomenon disappears in 5<$n$<20 but the $T_B$ >1 remains in 0<$n$<5. Combining Fig. 7(b) and Fig. 8(c,d), the phenomenon for $T_B$>1 appears near the source, and becomes remarkable in short diodes. Moreover, this phenomenon has little influence on the diode efficiency concerned in this paper.

## V. Experiments

Analyses above are made based on the model including smooth nonlinearity. The diode model containing clearance-nonlinearity and weak damping is also numerically analyzed. As shown in Fig. 9, the clearance-nonlinearity can repeat the bidirectional diode effect.

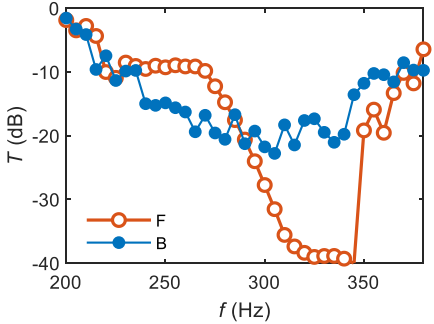

FIG. 9 Numerical results for the model containing clearance-nonlinearity. $A_0(f)$=60 μm and $c_0$=3×10$^{-5}$ s.

### A. Experimental methods

We performed experiments to demonstrate the performance of the bidirectional diode, as shown in Fig. 10. In the diode, the sphere and springs are made of steel; the hollowed parallelepiped and cylinder are made of aluminium alloy. We suspend the diode in a smooth guide rail through strings to reduce the friction between the diode and the rail. The left end of the diode is connected to an electromagnetic vibration exciter and the other end is fixed. A rubber is connected to the diode to reduce the boundary reflections. Three doppler laser vibrometers are used to measure the input and two output points, synchronously. The directly measured signal is the velocity, $v(t)$, and the displacement is $u(t)≈v(t)/2πf$ for the frequency $f$. The diving levels are tuned by the voltage of the amplifier for the electromagnetic excitor. Under a driving level, the power of the amplifier is constant such that the driving amplitude $A_0$ decreases as the central frequency $f$ increases, as shown in Fig. 11.

We still used sinusoidal wave packet as the input signal but the waveform is optimized in experiment. It is still a 10-periods of packet. Here, the middle 8 periods feature the constant amplitude $A_0$; the initial and the cut-off ends are smoothed to eliminate the impact wave induced by the jumping of amplitude. The peak amplitude of the transmitted packet in time domain is $A$, and the transmission denotes $T = A/A_0 =V/V_0$. As shown in Fig. 11, although there are influences of dispersion and subsequent vibrations when the incident packet is terminated, the peak value of the output packet can be clearly picked. Enough time interval, 0.5 s, between two packets is taken to make sure that the former packet will not influence the response of the next packet.

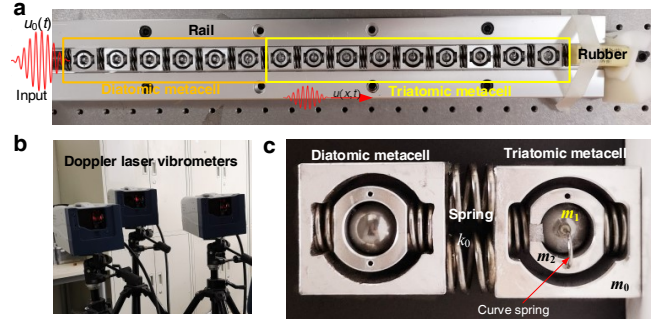

FIG. 10 (Color online) Experiments. (a) Experiment setup. (b) IThe doppler laser vibrometers used in experiment. (c) Pictures of the diatomic and triatomic metacells.

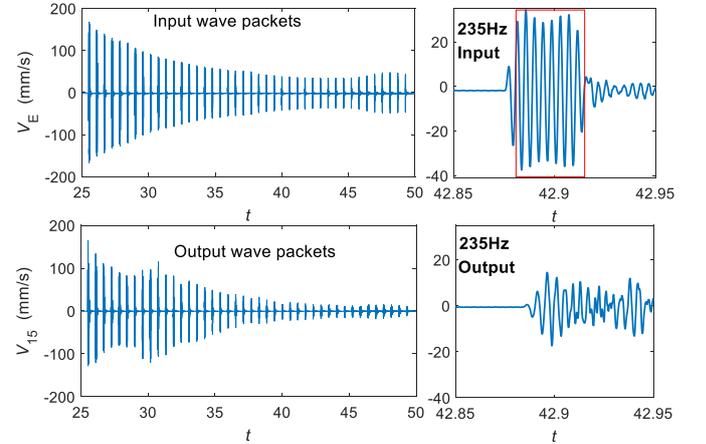

FIG. 11 Typical exciting and output signals, $V_E$ and $V_{15}$, at 5 V. Signals at 235 Hz are enlarged.

### B. Experimental demonstration

Experimental results are shown in Fig. 12. In the case with sinusoidal packet input, experiments under two excitation levels, 5, 9 V, are performed to give rise to the NLR bandgap in 200-275 Hz. The experimental trends in Fig. 12(a,b) agree very well with the numerical results in Fig. 9 and Fig. 2. $ΔT$ in LR and NLR respectively reaches -40 and 20 dB, which demonstrate the high-quality insulation.

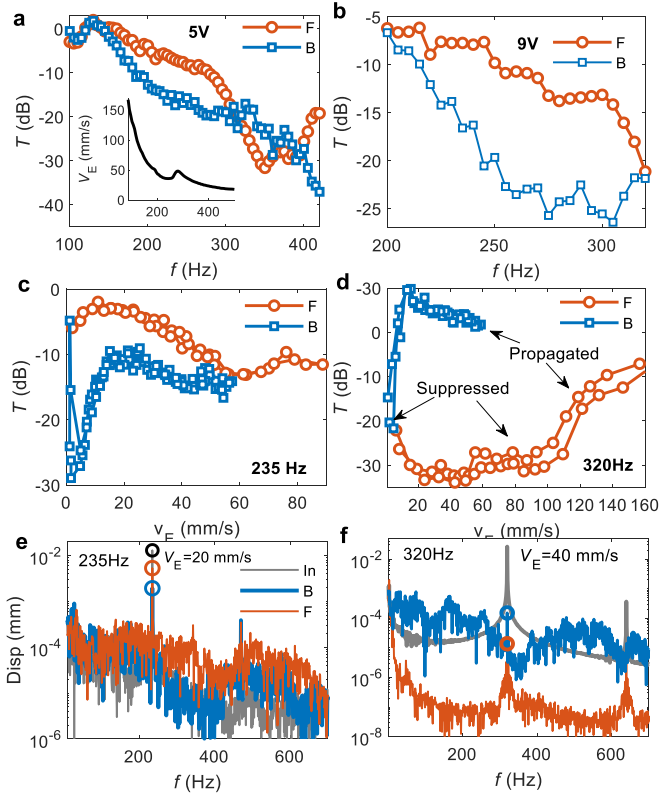

FIG. 12 (Color online) Experimental demonstrations. Inputs in (a, b,) are wave packets, in (c, d) are sweep-amplitude signals, and in (e, f) are sinusoidal signal with constant amplitude. (a) Results for level-5 V. (b) $T$ for level-9 V. The inset in (a) is the input velocity $V_E$ for level-5 V. Transmissions varying against $V_E$ at frequency (c) 235 and (d) 320 Hz. Two curves of each $T_F$ or $T_B$ correspond to ascending and descending processes. These transmissions are calculated in time domain. (e, f) Frequency spectra of the input forward and backward transmitted wave. Circles are the peak value at fundamental frequency.

Besides the sinusoidal packet, a sweep-amplitude signal at a given frequency (235 or 320 Hz) but with time-varying amplitude is also adopted in experiments, as shown in Fig. 13. Here, the transmission is calculated in different time interval:

$T(t, t+\Delta t)=20\log_{10}[A_{out}(t, t+\Delta t)/A_0(t, t+\Delta t)]$ (dB)

where $X(t, t+\Delta t)$ denotes the maximum value in the time interval $(t, t+\Delta t)$. As shown in Fig. 12(c,d), this experiment clearly describes the variation laws of the asymmetric transmission. We still show $T_{time}$ first. $\Delta T$ of the bidirectional diode is tunable by driving amplitude. When the incident velocity $V_E \to 0$, transmissions in both NLR and LR bands are symmetric due to linearity. Owing to the tiny clearance, small amplitude drives the metamaterial to generate enhanced nonlinearity, on which occasion reciprocities in both NLR and LR are broken. Positive $\Delta T$ at 235 Hz reaches 20 dB for a small amplitude $V_E \approx 5$ mm/s; after that, $T_B$ firstly increases and then decreases as $V_E$ ascends, but $T_F$ mainly behaves a descending trend because of the frequency shift of the peak $\Delta T$. However, the nonreciprocity at 235 Hz almost disappears near $V_E \approx 60$ mm/s ($A_0 \approx 40$ μm), which demonstrates the amplitude-dependent trends shown in Fig. 5.

In the LR bandgap represented by 320 Hz, the forward propagation remains suppressed when $V_E<110$ mm/s ($A_0<55$ μm), but $T_B$ jumps to a large value (-10 dB) at a small amplitude $V_E=13$ mm/s ($A_0 \approx 6.5$μm): A significant negative rectification $\Delta T \approx -40$ dB is achieved. Moreover, for $A_0>55$ μm, $T_F$ jumps to a large value and increases with $V_E$ on which occasion the forward energy starts to transfer.

Under typical amplitudes that can generate nonreciprocity, we analyze the spectra with constant sinusoidal inputs, as shown in Fig. 12(e,f). Figure 12(e) demonstrates that nonreciprocity in the NLR band preserves the frequency, and reciprocity of fundamental wave is also broken (see the peaks). In Fig. 12(f), the continuous spectrum of the backward wave confirms the chaotic property; although the fundamental wave is not the main component, reciprocity of fundamental wave is also broken. These results are consistent with theories.

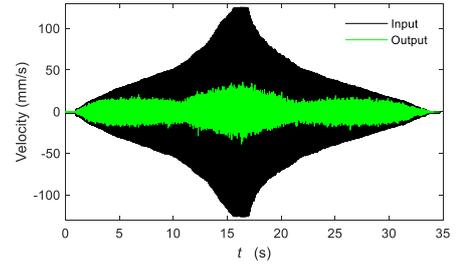

FIG. 13 Signals for sweep-amplitude experiment.

## VI. Conclusions

In conclusion, we demonstrated theoretically and experimentally the frequency-preserved, bidirectional, high-quality, low-frequency, small size (~one wavelength and can be sub-wavelength), and tunable elastic diode. This has been achieved by using elastic metamaterial with the intentional clearance inside metacells to create enhanced nonlinearity. Their transmission differences reach as much as 20 dB and -40 dB, respectively. We report three nonreciprocal mechanisms. The amplitude-dependent bandgap combining the interface reflection enables the frequency-preserved nonreciprocity of both the total energy and fundamental wave. The linear bandgap combining with the chaotic responses performs the monochromic–to–continuous nonreciprocity of total energy, while the propagation for fundamental wave is weakly nonreciprocal under weak damping. Increasing damping can break both types of reciprocity in the same band. Our study finds new physics, and paves ways to conceiving devices and metamaterials for asymmetric energy transmissions with reversible rectifying direction.

# Acknowledgement

XF and J-H Wen were funded by the National Natural Science Foundation of China (Projects No. 12002371, 11991032, and No. 11991034)

# Appendix

## A. Nonlinearity

In experiments, the nonlinear force $F_N(t)$ generated by the contact between the sphere and the hollow cylinder is a piecewise nonlinear function

$$F_N(t) = \begin{cases} k_1 u & \text{for } |u| \leq \delta_0 \\ k_1 u + A_c(u-\delta_0)^{3/2} & \text{for } u > \delta_0 \\ -k_1 u - A_c(-u-\delta_0)^{3/2} & \text{for } u < -\delta_0 \end{cases} \quad (A1)$$

Here, $\delta_0$ denotes the width of the clearance and $A_c = 2E_s\sqrt{r_s}/[3(1-v^2)]$, where $E_s$ and $v$ represent the elastic modulus and Poisson's ratio of the softer medium, respectively. $E_s$=70 GPa, $v$=0.3. The piecewise function $F_N(t)$ described by Eq. (A1) can be fitted with a smooth equation $k_1 p + k_N p^n$. The linear coefficient, $k_1$, is constant but the nonlinear coefficient, $k_N$, depends on the clearance, $\delta_0$. As shown in Fig. A1, by fitting the curve, one can use the cubic nonlinear equation to approximate the piecewise function, and $k_N \approx 3 \times 10^{12}$ N/m$^3$.

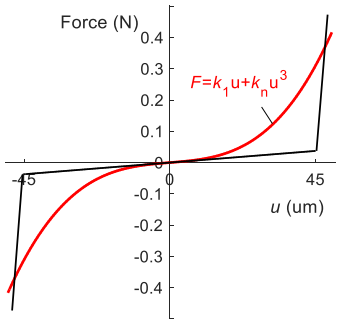

FIG. A1 The nonlinear forces-deformation curves. The black curve is the piecewise function.

## B. Dispersion theory

The dispersion relationship of the linear diatomic metamaterial is

$$\cos\mu = 1 - \frac{\omega^2 m_0}{2k} - \frac{\omega^2 \omega_r^2 m_r}{2k(\omega_r^2 - \omega^2)} \quad (A2)$$

where $k=k_0(1+i\omega c_0)$; $\mu=\kappa a$; $\kappa$ is the wave vector and $a$=27 mm is the lattice constant. The dispersion relationship of the linear triatomic metamaterial is

$$\cos\mu = 1 - \frac{m_0 \omega^2}{2k} - \frac{\omega^2 \omega_2^2 [m_1 \omega_1^2 + m_2(\omega_1^2 - \omega^2)]}{2k[m_2(\omega^2 - \omega_1^2)(\omega^2 - \omega_2^2) - m_1 \omega^2 \omega_1^2]} \quad (A3)$$

For the nonlinear triatomic metamaterial, we adopt the equivalent linearized approach based on the bifurcation of nonlinear local resonance to solve the dispersion effect. This approach is proposed in [37]. Similar with Eq. (A3), we have

$$\cos\mu = 1 - \frac{m_0 \omega^2}{2k} - \frac{\omega^2 \omega_2^2 [m_1 \omega_{1e}^2 + m_2(\omega_{1e}^2 - \omega^2)]}{2k[m_2(\omega^2 - \omega_{1e}^2)(\omega^2 - \omega_2^2) - m_1 \omega^2 \omega_{1e}^2]} \quad (A4)$$

Here, $\omega_{1e}$ is the equivalent nature frequency of oscillator $m_1$. It is solved with the bifurcation analyse. The nonlinear local resonance in the triatomic metacell can be solved with the first-order harmonic balance method. The equation is

$$\begin{cases} \omega^2 m_1 Y = k_1 P + 3k_N P^3/4 \\ (k_2 - \omega^2 m_2)(Y-P) - \omega^2 m_1 Y = k_2 A_0 \end{cases} \quad (A5)$$

This equation is obtained by specifying $u=A_0\sin\omega t$, $y=Y\sin\omega t$, $z=Z\sin\omega t$, $P=Y-Z$, in Eq. (1). For specified incident amplitude $A_0$, one obtains the saddle node bifurcation point of the first nonlinear resonance. The solution of this point is ($\omega_{J1}$, $P_{J1}$) (Please see Fig. 1c in Ref. [37]). The equivalent stiffness and nature frequency of the nonlinear oscillator $m_1$ is

$$k_{1e} = k_1 + 3k_N P_{J1}^2/4, \quad \omega_{1e} = \sqrt{k_{1e}/m_1} \quad (A6)$$

By specifying the frequency $\omega$, one can solve $\mu=\mu_R+i\mu_I$. The imaginary part $\mu_I$ characterize the wave attenuation. Moreover, based on Eqs. (A2) and (A4), we can calculate $\mu_I$ at every unit cell in the elastic diode. Then, we can obtain the forward and backward transmissions by using the attenuation $\exp[-\mu_I(n)]$. The reflection at the interface is not considered.

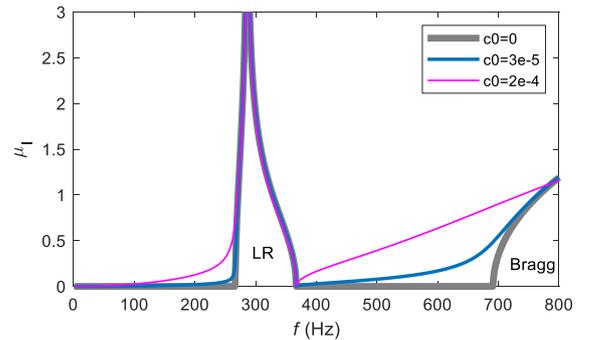

FIG. A2 Dispersion curve of the linear diatomic metamaterial.

# References

[1] M. O. Kolawole, *Electronics: from Classical to Quantum* (CRC Press, 2020).

[2] M. C. Gupta, and J. Ballato, *The Handbook of Photonics, 2nd Edition* (CRC Press, 2021).

[3] B. Li, L. Wang, and G. Casati, Thermal diode: Rectification of heat flux. Phys. Rev. Lett. **93**, 184301 (2004).

[4] N. Li, J. Ren, L. Wang, G. Zhang, P. Hänggi, and B. Li, Colloquium: Phononics: Manipulating heat flow with electronic analogs and beyond. Rev. Mod. Phys. **84**, 1045 (2012).

[5] G. Wehmeyer, T. Yabuki, C. Monachon, J. Wu, and C. Dames, Thermal diodes, regulators, and switches: Physical mechanisms and potential applications. Applied Physics Reviews **4**, 041304 (2017).

[6] V. F. Nesterenko, C. Daraio, E. B. Herbold, and S. Jin, Anomalous wave reflection at the interface of two strongly nonlinear granular media. Phys. Rev. Lett. **95**, 158702 (2005).

[7] B. Liang, X. S. Guo, J. Tu, D. Zhang, and J. C. Cheng, An acoustic rectifier. Nat. Mater. **9**, 989 (2010).

[8] H. Nassar, B. Yousefzadeh, R. Fleury, M. Ruzzene, A. Alù, C. Daraio, A. N. Norris, G. Huang, and M. R. Haberman, Nonreciprocity in acoustic and elastic materials. Nature reviews. Materials (2020).

[9] G. Wu, Y. Long, and J. Ren, Asymmetric nonlinear system is not sufficient for a nonreciprocal wave diode. Physical Review B **97**, 205423 (2018).

[10] A. A. Maznev, A. G. Every, and O. B. Wright, Reciprocity in reflection and transmission: What is a 'phonon diode'? Wave Motion **50**, 776 (2013).

[11] A. Souslov, B. C. van Zuiden, D. Bartolo, and V. Vitelli, Topological sound in active-liquid metamaterials. Nat. Phys. **13**, 1091 (2017).

[12] H. He, C. Qiu, L. Ye, X. Cai, X. Fan, M. Ke, F. Zhang, and Z. Liu, Topological negative refraction of surface acoustic waves in a Weyl phononic crystal. Nature **560**, 61 (2018).

[13] A. Y. Song, X. Q. Sun, A. Dutt, M. Minkov, C. Wojcik, H. Wang, I. Williamson, M. Orenstein, and S. Fan, PT-Symmetric Topological Edge-Gain Effect. Phys. Rev. Lett. **125**, 033603 (2020).

[14] H. Xue, Y. Yang, G. Liu, F. Gao, Y. Chong, and B. Zhang, Realization of an Acoustic Third-Order Topological Insulator. Phys. Rev. Lett. **122**, 244301 (2019).

[15] E. Riva, M. I. N. Rosa, and M. Ruzzene, Edge states and topological pumping in stiffness-modulated elastic plates. Physical Review B **101**, 094307 (2020).

[16] S. Li, I. Kim, S. Iwamoto, J. Zang, and J. Yang, Valley anisotropy in elastic metamaterials. Physical Review B **100**, 195102 (2019).

[17] M. Miniaci, R. K. Pal, B. Morvan, and M. Ruzzene, Experimental Observation of Topologically Protected Helical Edge Modes in Patterned Elastic Plates. Physical Review X **8**, 031074 (2018).

[18] M. I. N. Rosa, R. K. Pal, J. R. F. Arruda, and M. Ruzzene, Edge States and Topological Pumping in Spatially Modulated Elastic Lattices. Phys. Rev. Lett. **123**, 034301 (2019).

[19] Q. Wu, H. Chen, X. Li, and G. Huang, In-Plane Second-Order Topologically Protected States in Elastic Kagome Lattices. Physical Review Applied **14**, 014084 (2020).

[20] N. Boechler, G. Theocharis, and C. Daraio, Bifurcation-based acoustic switching and rectification. Nat. Mater. **10**, 665 (2011).

[21] J. Bunyan, K. J. Moore, A. Mojahed, M. D. Fronk, M. Leamy, S. Tawfick, and A. F. Vakakis, Acoustic nonreciprocity in a lattice incorporating nonlinearity, asymmetry, and internal scale hierarchy: Experimental study. Physical Review E **97**, 52211 (2018).

[22] I. Grinberg, A. F. Vakakis, and O. V. Gendelman, Acoustic diode: Wave non-reciprocity in nonlinearly coupled waveguides. Wave Motion **83**, 49 (2018).

[23] A. S. Gliozzi, M. Miniaci, A. O. Krushynska, B. Morvan, M. Scalerandi, N. M. Pugno, and F. Bosia, Proof of concept of a frequency-preserving and time-invariant metamaterial-based nonlinear acoustic diode. Sci. Rep.-UK **9**, 9560 (2019).

[24] C. Fu, B. Wang, T. Zhao, and C. Q. Chen, High efficiency and broadband acoustic diodes. Appl. Phys. Lett. **112**, 051902 (2018).

[25] Z. Li, B. Yuan, Y. Wang, G. Shui, C. Zhang, and Y. Wang, Diode behavior and nonreciprocal transmission in nonlinear elastic wave metamaterial. Mech. Mater. **133**, 85 (2019).

[26] C. Liu, Z. Du, Z. Sun, H. Gao, and X. Guo, Frequency-Preserved Acoustic Diode Model with High Forward-Power-Transmission Rate. Physical Review Applied **3**, 064014 (2015).

[27] J. Cui, T. Yang, and L. Chen, Frequency-preserved non-reciprocal acoustic propagation in a granular chain. Appl. Phys. Lett. **112**, 181904 (2018).

[28] H. Nassar, H. Chen, A. N. Norris, M. R. Haberman, and G. L. Huang, Non-reciprocal wave propagation in modulated elastic metamaterials. Proceedings of the Royal Society A: Mathematical, Physical and Engineering Sciences **473**, 20170188 (2017).

[29] G. Trainiti, and M. Ruzzene, Non-reciprocal elastic wave propagation in spatiotemporal periodic structures. New J. Phys. **18**, 83047 (2016).

[30] J. Huang, and X. Zhou, A time-varying mass metamaterial for non-reciprocal wave propagation. Int. J. Solids Struct. **164**, 25 (2019).

[31] P. A. Deymier, V. Gole, P. Lucas, J. O. Vasseur, and K. Runge, Tailoring phonon band structures with broken symmetry by shaping spatiotemporal modulations of stiffness in a one-dimensional elastic waveguide. Physical Review B **96**, 064304 (2017).

[32] Y. Chen, X. Li, H. Nassar, A. N. Norris, C. Daraio, and G. Huang, Nonreciprocal Wave Propagation in a Continuum-Based

Metamaterial with Space-Time Modulated Resonators. Physical Review Applied **11**, 064052 (2019).

[33] Z. Liu, X. Zhang, Y. Mao, Y. Y. Zhu, Z. Yang, C. T. Chan, and P. Sheng, Locally resonant sonic materials. Science **289**, 1734 (2000).

[34] Y. Wang, H. Zhao, H. Yang, J. Zhong, D. Zhao, Z. Lu, and J. Wen, A tunable sound-absorbing metamaterial based on coiled-up space. J. Appl. Phys. **123**, 185109 (2018).

[35] X. Fang, J. Wen, B. Bonello, J. Yin, and D. Yu, Ultra-low and ultra-broad-band nonlinear acoustic metamaterials. Nat. Commun. **8**, 1288 (2017).

[36] O. R. Bilal, A. Foehr, and C. Daraio, Bistable metamaterial for switching and cascading elastic vibrations. Proceedings of the National Academy of Sciences **114**, 4603 (2017).

[37] X. Fang, J. Wen, H. Benisty, and D. Yu, Ultrabroad acoustical limiting in nonlinear metamaterials due to adaptive-broadening band-gap effect. Physical Review B **101**, 104304 (2020).

[38] X. Fang, J. Wen, J. Yin, D. Yu, and Y. Xiao, Broadband and tunable one-dimensional strongly nonlinear acoustic metamaterials: Theoretical study. Physical Review E **94**, 052206 (2016).

[39] X. Fang, J. Wen, B. Bonello, J. Yin, and D. Yu, Wave propagation in one-dimensional nonlinear acoustic metamaterials. New J. Phys. **19**, 053007 (2017).

# Supplementary Material for

# Bidirectional elastic diode with frequency-preserved nonreciprocity

Xin Fang[1,2], Jihong Wen[1], Li Cheng[2], and Baowen Li[3]

[1] Laboratory of Science and Technology on Integrated Logistics Support, College of Intelligent Science, National University of Defense Technology, Changsha, Hunan 410073, China
[2] Department of Mechanical Engineering, Hong Kong Polytechnic University, Hong Kong, China.
[3] Paul M Rady Department of Mechanical Engineering and Department of Physics, University of Colorado, Boulder, Colorado 80309, USA

## Note 1. Perfect match layer

Many literatures show that asymmetric linear model can also realize the asymmetric wave propagation. However, in linear model, the asymmetric transmission relies on controlling the boundary conditions instead of breaking reciprocity. To rigorously demonstrate nonreciprocity in our elastic diode in theory, we must carefully check the boundary conditions applied in numerical models. The condition should result in perfect symmetric transmission in the linear model. For this purpose, two perfect match layers (PMLs) are connected to the two ends of the "diode", respectively. The PML plays a "buffer" role to dissipate the incident energy, i.e., it models the non-reflecting boundary.

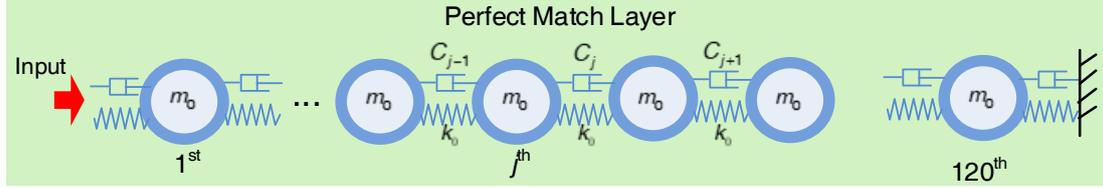

Fig. S1 Model of the perfect match layer

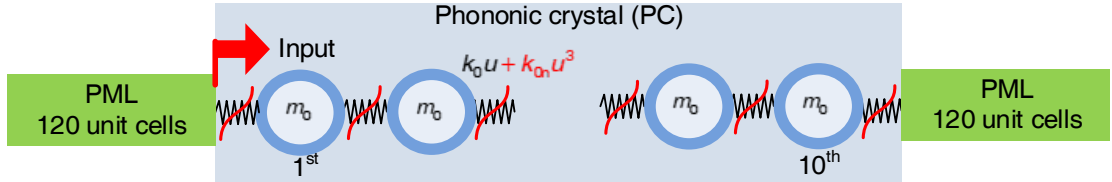

Fig. S2 The model used to test the PML

As shown in Fig. S1, the PML is a linear monoatomic chain. It contains 120 cells. The incident wave first arrives at the 1st unit cell. The motion equation of the 1~119th unit cell is

$$m_0\ddot{q}_j = k_0(q_{j+1} + q_{j-1} - 2q_j) + C_j(\dot{q}_{j+1} - \dot{q}_j) - C_{j-1}(\dot{q}_j - \dot{q}_{j-1}) \qquad (1)$$

The last unit cell (120th) is connected to a fixed boundary, and its motion equation is

$$m_0\ddot{q}_{120} = k_0(q_{119} - 2q_{120}) - C_{120}\dot{q}_{120} - C_{119}(\dot{q}_{120} - \dot{q}_{119}) \qquad (2)$$

where $q$ denotes the displacement; $m_0$ and $k_0$ are the primary mass and stiffness, and their values are same with those in the triatomic or diatomic metacell in the diode; $C_j$ denotes the damping coefficient. The damping coefficient is constant in the 1st–100th unit cell, and $C_j = 0.0015k_0$. $C_j = 100\sqrt{(j-100)/20}$, $j = 101\cdots 120$. Moreover, to obtain the rigorous symmetric transmission in linear model, one should also consider the way for inputting a signal, $u_0$, into the system. The motion equation of the 1st metacell in the diode (not the PML) is

$$m_0\ddot{u}_1 = k_0(u_2 + u_0 + q_1 - 2u_1) + c_0 k_0(\dot{u}_2 - \dot{u}_1) + k_2(Q_1 - u_1) \tag{3}$$

where $Q_1 = z_1$ or $r_1$, which depends on the incident directions, forward or backward; $q_1$ is the displacement of the 1st oscillator in PML. For the first unit cell in the PML, the motion equation is

$$m_0\ddot{q}_1 = k_0(q_2 + u_X - 2q_1) + 0.0015 k_0(\dot{q}_2 + \dot{u}_X - 2\dot{q}_1) \tag{4}$$

Here $u_X$ is the displacement of the oscillator in the diode that connects with the PML. We note that **the wave attenuation in PML is attributed to damping. Therefore, the PML is effective not only for the fundamental wave in linear cases, but also for the multiple harmonics in the nonlinear cases.**

To demonstrate the effectiveness of the PML, we establish a model shown in Fig. S2. Here, a nonlinear phononic crystal (PC) is inserted between two PMLs. The PC consists of 10 monoatomic nonlinear oscillators. Here we consider cubic nonlinearity. The motion equation of the $j^{th}$ nonlinear oscillator is

$$m_0\ddot{u}_j = k_0(u_{j+1} + u_{j-1} - 2u_j) + k_{0n}(u_{j+1} + u_{j-1} - 2u_j)^3 \tag{5}$$

The nonlinear coefficient is $k_{0n}$=3e13 N/m³ for the nonlinear case. For linear case, $k_{0n}$=0.

Typical wave propagations are shown in Fig. S3, Fig. S4, and Fig. S5. For the case of 200 Hz in Fig. S3, when inputting the sinusoidal wave packet, the PML can perfectly absorb the incident wave without reflecting it back to the PC. **In the nonlinear model, although 3rd, 5th, 7th harmonics are generated, the PML is still effective.**

However, the PML is a physical model to simulate the infinite boundary. As well known, it cannot always remain "perfect" in all conditions. It is used to **reduce** the numerical errors instead of **eliminating** numerical errors. For example, in Fig. S4, although it is a linear model, the output wave at 80 Hz is amplified. This is the standing wave effect. It happens because the PML cannot absorb all wave energy very quickly if the frequency is low. Moreover, when inputting a standard single-frequency sinusoidal wave, as shown in Fig. S5, there is always a very low frequency wave response due to the vibration of the whole model (including PC and 2 PMLs). In this case, the standing wave effect may be more significant in our diode model due to the reflection between different boundaries.

Moreover, we note that the monoatomic chain used to simulate the PML is a dispersion model. This PML is available only when the incident frequency is lower than the bandgap frequency $f_{st} = \sqrt{k_0/m_0}/\pi$ of the PML. In this paper, $f_{st}$ =664 Hz. For example, as shown in Fig. S4b, the 400 Hz output wave packet is reduced due to the dispersion effect because the spectrum of a wave packet is broadband.

Fortunately, the standing wave and dispersion effects do not change our conclusions for reasons: (1) We mainly focus on the properties between 150-400 Hz where the influences of standing waves are small; (2) The PML is still very effective in these frequency ranges, which minimizes the influence of reflections at free or fixed end (Note: Without the PML, one end of the model is free or fixed, which induces significant resonances); (3) We use the signal procession methods in Note 2 to reduce numerical errors.

In short, many measures are taken to minimize numerical errors and make our conclusions be rigorous. Actually, these measures are very successful. Another evidence is shown in Fig. 1f in the main text: The forward and backward transmissions are rigorously equal in the linear model.

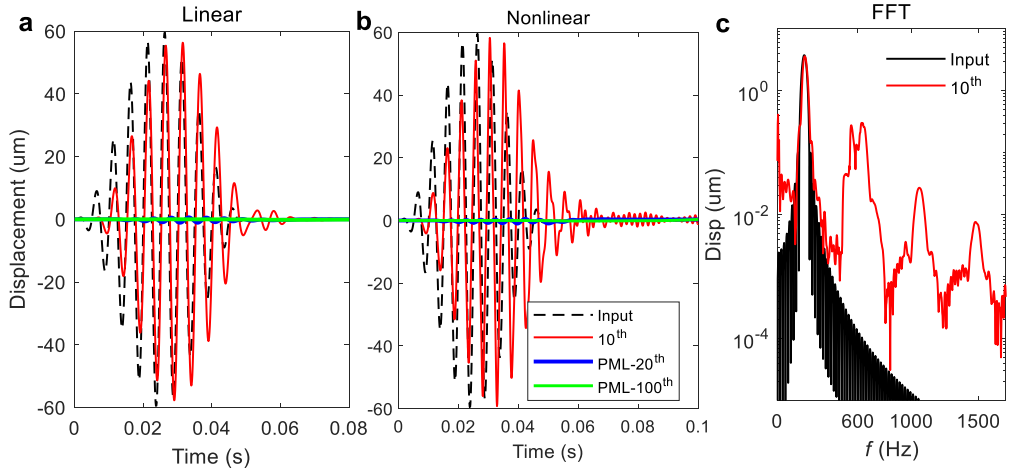

Fig. S3 Wave propagation in the testing model shown in Fig. S2. The central frequency of the incident wave packet is 200 Hz. **a**, Linear case. **b**, Nonlinear case. **c**, The frequency spectra of the input and output waves. The legends in **a** and **b** are same: "10th" represents the 10th unit cell; "PML-$n$th" represent the nth unit cell in the PML. For the nonlinear case, $A_0$=60 μm.

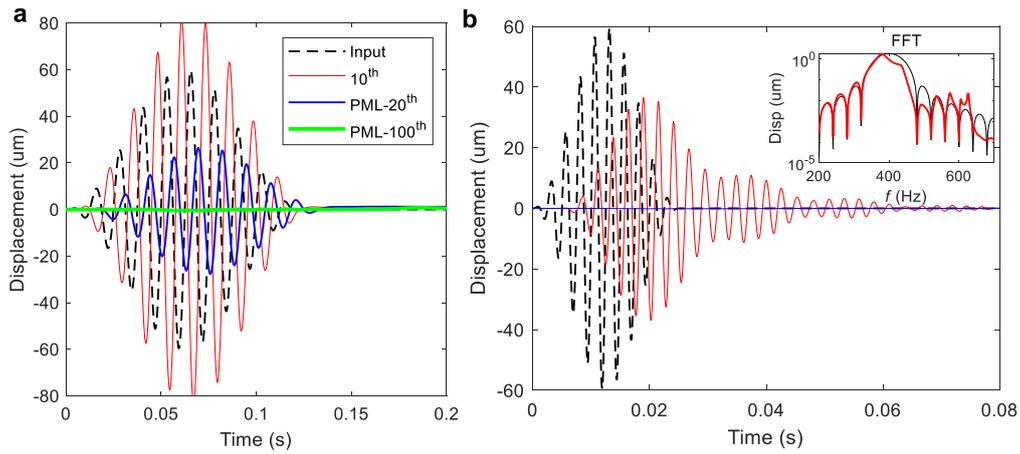

Fig. S4 Wave propagation in the testing model shown in Fig. S2. The central frequencies of the incident wave packet are 80 Hz in panel a and 400 Hz in panel b. Only linear case is considered here.

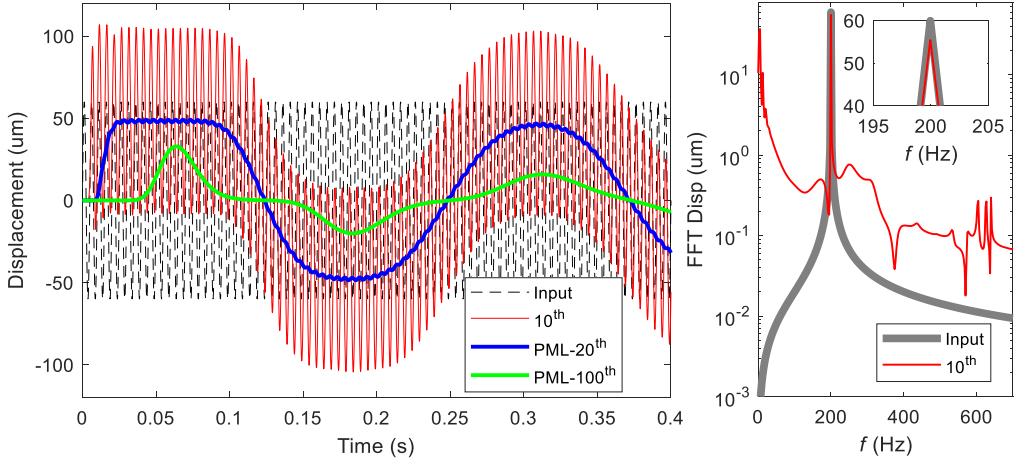

Fig. S5 Wave propagation in the testing model shown in Fig. S2. Here the incident wave is a standard sinusoidal signal whose frequency is 200 Hz. The right panel shows the frequency spectra. Only linear case is considered here.

# Note 2. Signal processing methods

When inputting the 10-periods of wave packets, the total energy of propagation wave is characterized by the maximum amplitude of the packet at every unit cell:

$$A_n^{(\text{time})} = \max[u_n(t)] \quad \text{for} \quad 0 < t < 10/f$$

Here $f$ denotes the central frequency of the packet. This equation means that any reflecting waves from the fixed end (see Fig. S1) are not considered.

When inputting the standard sinusoidal wave, the simulation time is 1.0 second. The period is $T=1/f$. To exclude the influence of the very low-frequency vibration shown in Fig. S5, three steps are taken to calculate the real amplitude of the fundamental wave in time domain.

Step 1: Divide the signal into $J$ segments, and the length of every segment is $1.5T$.

Step 2: Extract the maximum and minimum values, $U_{\max,j}$ and $U_{\min,j}$, in each segment.

Step 3: Calculate the amplitude in time domain:

$$A_n^{(\text{time})} = (\sum_{j=1}^{J} U_{\max,j} - \sum_{j=1}^{J} U_{\min,j})/2$$